\documentclass[pdflatex,sn-mathphys-num]{sn-jnl}

\usepackage{graphicx}%
\usepackage{multirow}%
\usepackage{amsmath,amssymb,amsfonts}%
\usepackage{amsthm}%
\usepackage{mathrsfs}%
\usepackage[title]{appendix}%
\usepackage{xcolor}%
\usepackage{textcomp}%
\usepackage{manyfoot}%
\usepackage{booktabs}%
\usepackage{algorithm}%
\usepackage{algorithmicx}%
\usepackage{algpseudocode}%
\usepackage{listings}%
\usepackage{listings}
\usepackage{color}
\usepackage{xcolor}
\usepackage{float}
\usepackage{physics}

\usepackage{amsthm}

\definecolor{dkblue}{rgb}{0,0.39,0}
\definecolor{gray}{rgb}{0.66,0.66,0.66}
\definecolor{mauve}{rgb}{0.91,0.33,0.50}
\definecolor{gold}{rgb}{1,0.84,0}

\begin{document}

\title[Predicting Water Quality using Quantum Machine Learning: The Case of the Umgeni Catchment (U20A) Study Region]{Predicting Water Quality using Quantum Machine Learning: The Case of the Umgeni Catchment (U20A) Study Region}

\author*[1]{\fnm{Muhammad Al-Zafar} \sur{Khan}}\email{Muhammad.Al-ZafarKhan@zu.ac.ae}

\author[1,2]{\fnm{Jamal} \sur{Al-Karaki}}\email{Jamal.Al-Karaki@zu.ac.ae}

\author[3]{\fnm{Marwan} \sur{Omar}}\email{momar3@iit.edu}



\affil*[1]{\orgname{College of Interdisciplinary Studies (CIS), Zayed University}, \state{Abu Dhabi}, \country{UAE}}

\affil[2]{\orgdiv{College of Engineering}, \orgname{The Hashemite University}, \city{Zarqa}, \country{Jordan}}

\affil[3]{\orgdiv{Faculty of Information Technology Management}, \orgname{Illinois Institute of Technology}, \city{Chicago}, \country{USA}}


\abstract{In this study, we consider a real-world application of QML techniques to study water quality in the U20A region in Durban, South Africa. Specifically, we applied the quantum support vector classifier (QSVC) and quantum neural network (QNN), and we showed that the QSVC is easier to implement and yields a higher accuracy. The QSVC models were applied for three kernels: Linear, polynomial, and radial basis function (RBF), and it was shown that the polynomial and RBF kernels had exactly the same performance. The QNN model was applied using different optimizers, learning rates, noise on the circuit components, and weight initializations were considered, but the QNN persistently ran into the dead neuron problem. Thus, the QNN was compared only by accraucy and loss, and it was shown that with the Adam optimizer, the model has the best performance, however, still less than the QSVC.}



\maketitle
\section{Introduction}
Water quality assessment remains a critical challenge in environmental monitoring and public health protection. Traditional machine learning (ML) approaches have demonstrated success in predicting water quality parameters \cite{haghiabi2018water,ahmed2019machine,zhu2022review,tung2020survey,azrour2022machine,talukdar2023predicting,shamsuddin2022water,rustam2022artificial}, yet they often struggle with the complex, nonlinear relationships inherent in aquatic systems. Quantum Machine Learning (QML) offers promising advantages through its ability to efficiently process high-dimensional data and exploit quantum phenomena like superposition and entanglement. The utility of QML has been demonstrated in various applications related to financial fraud detection \cite{innan2024financial,innan2024financial2,kyriienko2022unsupervised,wang2022integrating,liu2018quantum}, drug discovery \cite{batra2021quantum,mensa2023quantum,li2021drug,bhatia2023quantum}, chemistry and medical applications \cite{atz2022delta,huang2020quantum,xia2018quantum,innan2023electronic}, materials science \cite{naseri2023quantum,lourencco2024exploring,vedavyasa2024classification}, and many other areas.

This research explores the application of QML algorithms to water quality prediction, specifically implementing the quantum support vector classifier (QSVC) and quantum neural networks (QNNs). We demonstrate that quantum approaches can capture subtle correlations between water quality indicators, including the presence of various chemical compounds in the water ($\text{NH}_{3},\text{NO}_{2},\text{NO}_{3},\text{SO}_{4}$ and others), as well as sediment trappings, flow rates, flood attenuation, chemical assimilation (phosphates, nitrates, toxicants), and other factors like turbidity. Our methodology leverages both classical preprocessing techniques and quantum feature maps to encode water quality data into quantum states, followed by measurement-based prediction. The results indicate that QML models can achieve comparable or superior accuracy to classical methods while requiring fewer training parameters and computational resources. Many studies enlist the potential advantages that QML has over classical ML, so we choose to avoid a restatement here. However, we should remember that we strongly believe that QML is an experimental science. Therefore, the advantages of QML over classical ML can only be realized once a particular application is explored, as we have seen in our study.

We consider the real-world study area from where this data was collected: The City of Durban, situated in KwaZulu-Natal on the east coast of South Africa, has a rich and currently developing history of Quantum Computing, arguably being the first city in South Africa that envisioned the potential of QC and its applications. Possibly, the doyen of QC in South Africa, Francesco Petruccione, introduced this field via the Quantum Research Group at the University of KwaZulu-Natal, and this was further solidified when, together with his then student, Maria Schuld, they co-authored the famous two-part magnum opus on Quantum Machine Learning (QML) \cite{schuld2018supervised,schuld2021machine}. Thus, it is almost fairytale-like that we use a technology pioneered in this city to study the quality of water in the city. 

In recent years, the quality of water in the Durban area has supposedly been in the decline due to various factors, including accusations of illegal dumping of sewerage into the beaches, amongst others, therefore affecting 
sanitation, consumption, tourism -- a major source of income -- in the city and the province \cite{bond2011durban,bond2019tokenistic,nkosi2014analysis}. 

The foremost objective of this research is to approach the issue of water quality from a non-political, unbiased, physical sciences perspective and apply the techniques of QML to predict whether water quality is good or bad for use. The term \textit{use} is used broadly to mean leisurely activities such as swimming. 
 
However, we acknowledge that the methods used here are not novel. Since 2018, there has been a plethora of applications for these methods in the literature, and the application domain is certainly unique. The use of real-world data collected by field agents of a credible organization certainly serves as an original contribution to the field of applications of QML.

This paper is divided as follows:

In Sec.\ref{related work}, we review some related works that are congruous to our study.

In Sec.\ref{theory}, we provide the theory for the models used. 

In Sec.\ref{results}, we give the results of the model that were used in the form of metrics and discuss their implications. 

In Sec.\ref{conclusion}, we provide conclusive remarks on this paper whereby we reflect upon our findings and give directions for future research. 


\section{Related Work}\label{related work}

To our knowledge, there exists no other study that directly applies QML for the study of water quality, thus, our claim is that the application is completely new. Therefore, no direct literature applies to our study, but analogous studies that collected similar data to study weather phenomena and considered applications in agriculture, amongst others.

In \cite{grzesiak2024flood}, the authors use the Wupper River in Germany as a study area to build QML models to better understand flood forecasting during the period of 2023. By comparing classical and quantum ML models, the authors try to show the benefit of using QML. While this paper does not technically predict water quality, it shares some parallels with our study in the sense of using numerical variables related to water and choosing a demarcated study area. Similarly, in \cite{lin2024quantum}, the authors apply a novel QML model that combines long short-term memory (LSTM) models with QNNs to produce Quantum Trains (QT) and study flood prediction as an application. The novelty of their method lies in the manner in which the QT model reduces the number of trainable parameters. 

In \cite{basit2025optimizing}, the authors explore the applications of QSVCs and QNNs to enhance crop yields. By using data related to water quality and soil collected over a long period, the authors could show that QML has the potential to improve crop yields, and the QML models showed superior performance to classical methods. Similarly, in \cite{setiadi2024rice}, another application of QML to crop yield, in this case rice, is considered. The model was highly accurate regarding the MSE, MAE, and coefficient of determination. 

There are several other application-based studies that explored the aforementioned areas, however, in essence QML techniques were applied to data and the quantum advantage was demonstrated by reporting high metrics.


\section{Theory}\label{theory}
In this section, we present the theories underlying the models we use in this study.  

\subsection{Quantum Support Vector Classifiers}
\begin{figure}[H]
    \centering
    \includegraphics[width=1\linewidth]{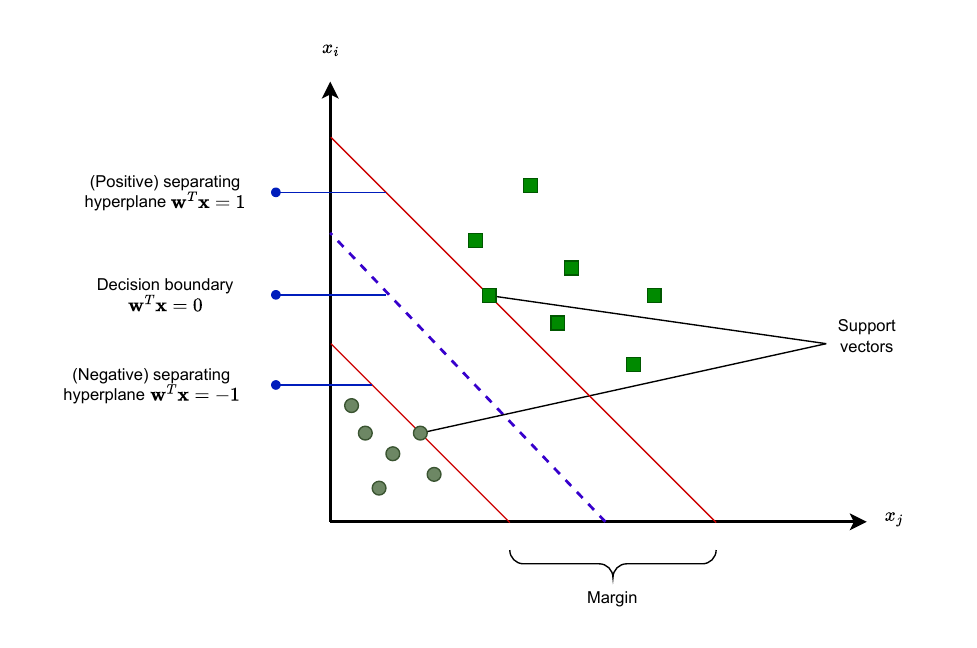}
    \caption{Diagrammatic depiction of the operation of an SVM. The goal is to maximize the margin between the decision boundary and the support vectors while minimizing misclassifications.}
    \label{fig2}
\end{figure}

Support Vector Machines (SVMs) work by finding the optimal separating hyperplane that segregates data points of different classes with the maximum margin. For a dataset $\mathcal{D}=\left\{\mathbf{x}_{i},y_{i}\right\}_{i=1}^{n}$ with $\mathbf{x}=\left(x_{1},x_{2},\ldots,x_{n}\right)$ being the features set, and $y$ being the predictor variable, we aim to solve the optimization problem
\begin{equation}\label{svm standard form}
\begin{aligned}
&\underset{\mathbf{w}}{\min}\left(\frac{1}{2}||\mathbf{w}||^{2}+C\sum_{i=1}^{n}\xi_{i}\right) \\
&\text{s.t.}\;y_{i}\left(\mathbf{w}^{T}\phi(x_{i})+b\right)\geq 1-\xi_{i}, \quad\xi_{i}\geq 0,
\end{aligned}
\end{equation}
where $\mathbf{w}$ are the weights, $C\in\left[0,1\right]$ is the control parameter, $\xi_{i}$ are the slack variables for misclassifications, and $\phi$ is the kernel which maps the features to the higher-dimensional feature space. In Fig.\ref{fig2}, we provide a two-dimensional rendition of the operation of an SVM. By introducing the Lagrange multiplier $\alpha_{i}$ we re-state Eqn.\eqref{svm standard form} in the form
\begin{equation}
\begin{aligned}
&\underset{\alpha}{\max}\left(\sum_{i=1}^{n}\alpha_{i}-\frac{1}{2}\sum_{i=1}^{n}\sum_{j=1}^{n}\alpha_{i}\alpha_{j}y_{i}y_{j}K(x_{i},x_{j})\right) \\
&\text{s.t.}\;\sum_{i=1}^{n}\alpha_{i}y_{i}=0, \quad 0\leq\alpha_{i}\leq C.
\end{aligned}
\end{equation}
The feature map $\phi(x)$, which takes the data to a high-dimensional space using the kernel
\begin{equation}
K(x_{i},x_{j})=
\begin{cases}
x_{i}^{T}x_{j},\text{for linear kernels}, \\
\left(\beta x_{i}^{T}x_{j}+r\right)^{D}, \text{for polynomial kernels}, \\
\exp\left(-\beta||x_{i}-x_{j}||^{2}\right),\text{for radial basis function (RBF) kernels},
\end{cases}
\end{equation}
where $\beta$ is the scaling parameter, $r$ is the coefficient, and $D$ is the polynomial degree.  

Specifically, classical data $x_{i}$ is mapped into quantum states $\ket{\psi(x_{i})}$ using a feature map scheme. Thereafter, the quantum kernel
\begin{equation}
K(x_{i},x_{j})=|\bra{\psi(x_{i})}\ket{\psi(x_{j})}|^{2},
\end{equation}
is calculated. The purpose of the quantum kernel is to measure the similarity between pairs of quantum states. The feature map is implemented using a VQC with quantum gates. By running the quantum circuit and measuring the outcomes, $K$ is calculated. Thereafter, the classical SVM method is applied to find the optimal hyperplane that separates the quantum states in the high-dimensional feature space, and data points are classified accordingly. 

\subsection{Quantum Neural Networks}
Quantum Neural Networks (QNNs) are the QML analogs of classical NNs that leverage quantum properties to process data. They have been shown to be particularly well suited for tasks in ML, optimization, and signal processing, with potential advantages in speed and expressive power over classical counterparts.

In practice, a typical QNN consists of the following architecture components:
\begin{enumerate}
\item \textbf{Classical-to-Quantum Data Encoding:} Classical data is encoded into a quantum state. Mathematically, a classical datapoint $x_{i}$ is encoded into a quantum state $\ket{\psi(x_{i})}$. The commonly used encoding schemes are 
\begin{equation}
\ket{\psi}= 
\begin{cases}
\sum_{i=1}^{2^{n}-1}x_{i}\ket{i},\text{with}\;\sum_{i}|x_{i}|^{2}=1,\text{amplitude encoding}, \\
\bigotimes_{i=1}^{n}\left(\cos x_{i}\ket{0}+\sin x_{i}\ket{1}\right),\text{angle encoding}. 
\end{cases}
\end{equation}
\item \textbf{Parametrized/Variational Quantum Circuit (PQC/VQC):} A sequence of quantum gates with trainable parameters analogous to weights in classical networks. The VQC is the beating heart of the QNN and is composed of a collection of quantum gate operations such as rotations and control gates. For example, for rotations $R_{\nu}(\theta)=\exp(-\imath\theta\nu/2)$. The VQC is then a matrix that operates on the input quantum states according to
\begin{equation}
\ket{\psi_{\text{out}}}=U(\boldsymbol{\theta})\ket{\psi_{\text{in}}},
\end{equation}
for trainable parameters $\boldsymbol{\theta}$.
\item \textbf{Measurement:} Observables of the quantum state are measured to produce outputs. For an observable $\hat{O}$, measurement is represented by
\begin{equation}
\langle\hat{O}\rangle=\bra{\psi_{\text{out}}}\hat{O}\ket{\psi_{\text{in}}}.   
\end{equation}
\item \textbf{Loss Function:} A classical loss function evaluates the difference between predicted and target values. For a binary classification problem with targets $y\in\left\{0,1\right\}$, the mean-squared error loss is used
\begin{equation}
\mathcal{L}(\boldsymbol{\theta})=\frac{1}{n}\sum_{i=1}^{n}\left[f(x_{i},\boldsymbol{\theta})-y_{i}\right]^{2},
\end{equation}
where $f(x_{i},\boldsymbol{\theta})$ is the output produced by the QNN for input $x_{i}$. 
\item \textbf{Parameter Optimization:} Gradients of the loss function are computed to update the parameters. The goal is to minimize the loss function by updating the parameters $\boldsymbol{\theta}$. 
\end{enumerate}

\begin{figure}[H]
    \centering
    \includegraphics[width=1.1\linewidth]{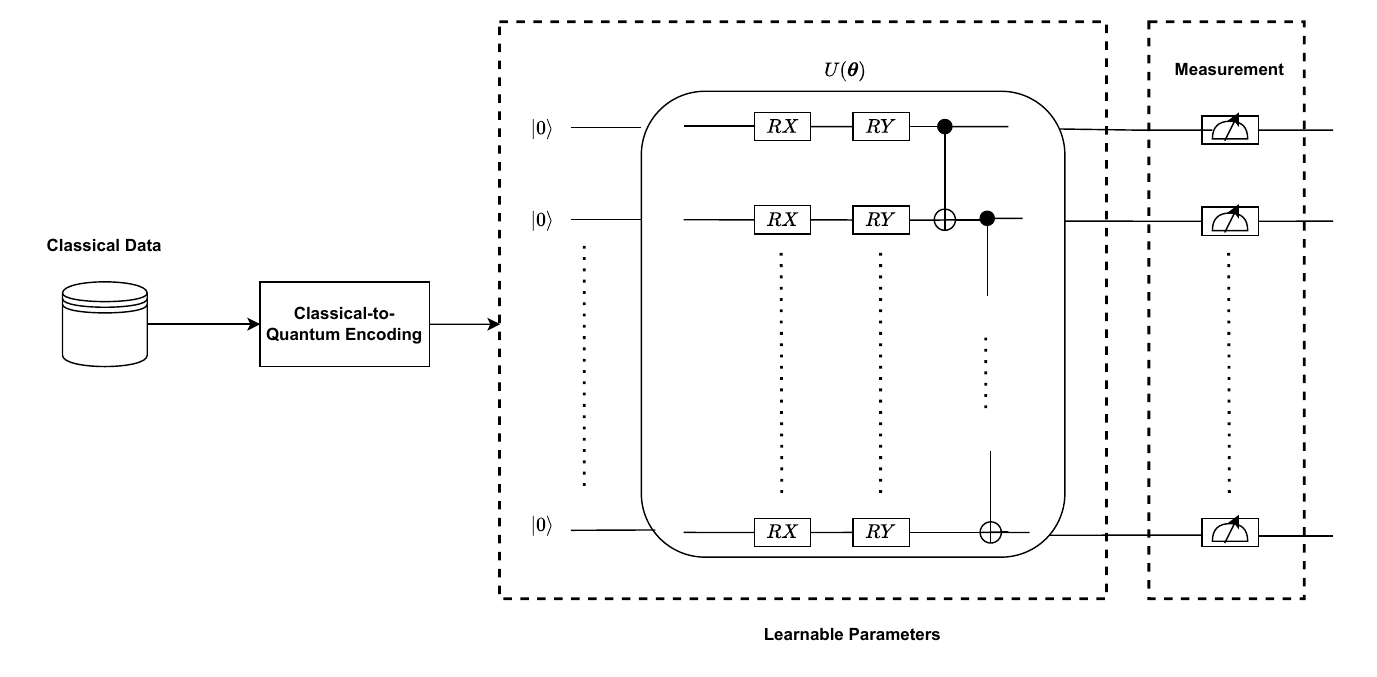}
    \caption{Architecture of a QNN. Input data is encoded into quantum states via a feature map. Thereafter, the state is initialized to the ground state $\ket{0}$ and fed into the unitary layer, which applies an ansatz to learn the parameters $\boldsymbol{\theta}$, and measurement is carried out. The process is repeated until a loss function $\mathcal{L}$ is sufficiently minimized.}
    \label{fig3}
\end{figure}

In Fig.\ref{fig3}, we provide a diagrammatic depiction of the architecture of a typical QNN and describe its operation, 

\section{Results}\label{results}
\begin{figure}[H]
    \centering
    \includegraphics[width=1.0\linewidth]{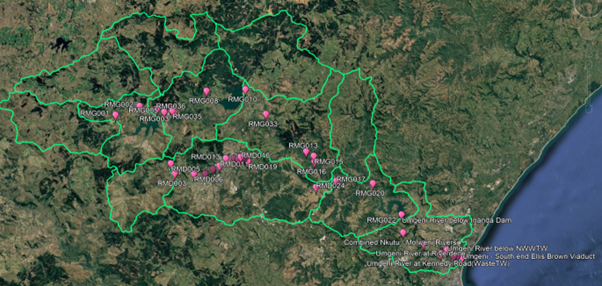}
    \caption{Map demarcating the U20A study area.}
    \label{fig1}
\end{figure}

The data comprised 32 data points representing 32 locations on the map where these variables were measured. The feature \textbf{E.coli - (MPN/100mL)} measures the amount of Escherichia coli bacteria in the water at that measurement point. If the amount was at most 235 MPN/100 mL then it was acceptable, if it exceeded this threshold it was regarded as unacceptable. Using this methodology, a predictor variable \textbf{Acceptable / Not Acceptable (For Recreation)} was created. In Fig.\ref{fig1}, we depict the study region considered. 

Within this criteria, only 3 locations had acceptable water quality while 29 had unacceptable water quality. Thus, the predictor variable was severely out of balance. Thus, class balancing was done using random oversampling. 

\subsection{QSVC}

\begin{table}[h]
\centering
\begin{tabular}{lllllll}
\hline
\textbf{Kernel} &\textbf{Accuracy} &\textbf{F1} &\textbf{Precision} &\textbf{Recall} &\textbf{AUROC} &\textbf{AUPRC} \\
\hline 
Linear &0.5833 &0.7059 &0.5455 &1.0000 &0.5833 &0.5455 \\
Poly &0.7500 &0.8000 &0.6667 &1.000 &0.7500 &0.6667 \\
RBF &0.7500 &0.8000 &0.6667 &1.000 &0.7500 &0.6667 \\
\hline 
\end{tabular}
\caption{Model metrics for the QSVC.}
\label{tab1}
\end{table}

From Tab.\ref{tab1}, we see that while with the linear kernel, the model achieves a perfect recall (1.0000), its accuracy, precision, and other metrics are relatively lower. This suggests that the linear kernel might not be the best choice for this specific dataset. Both kernels perform similarly well, with high accuracy, F1-score, precision, recall, AUROC, and AUPRC. This indicates that these kernels are better suited for capturing the underlying patterns in the data. The reason why the poly and RBF kernels produce exactly the same metrics may be attributed to several factors, including linear separability in the data. However, we see that we get a lower accuracy with the choice of a linear kernel; secondly, it could indicate that the model hyperparameters for both or either of the models are not properly fine-tuned, i.e., for the poly kernel, the choice of the degree of 1 might be too simplistic. For the RBF kernel, the choice of the complexity of the decision boundary $\gamma\to 0$ might not be optimal. Of course, one may experiment with different parameter values and obtain the optimal choice, but this is beyond the scope of our research objectives in this paper.   

\subsection{QNN}
Upon the initial training of the QNN, the model ran into the ``dead neuron problem'', i.e., all neurons were giving the same output irrespective of the input. After 50 epochs of training, the model continuously gave a loss of 0.4996. This problem is attributed to two issues
\begin{enumerate}
\item The ReLU function in the intermediate layers kept ``dying'' potentially due to a too high learning rate selected. 
\item The network cannot learn due to exploding/vanishing gradients.
\end{enumerate}

In the next iteration of model training, the architecture was adjusted and the abovementioned issues were addressed. Specifically, the learning rate was adjusted from 0.1 to 0.001, and the weight initialization was changed to the Xavier initialization 
\begin{equation}
\mathbf{w}\sim\mathcal{U}(-x,x),\quad x=\sqrt{\frac{6}{\text{fan}_{\text{in}}+\text{fan}_{\text{out}}}}, 
\end{equation}
where $\mathcal{U}$ is a uniform distribution, $\text{fan}_{\text{in}}$ is the number of input units to a layer, and $\text{fan}_{\text{out}}$ is the number of output units to a layer. 

Despite these adjustments, the QNN model still gave a constant loss throughout all epochs. Lastly, a noisy QNN model was trained using depolarization noise and amplitude damping, and the same result was obtained. Thus, it does not make sense to compare model metrics like F1, precision, and recall because they all had a constant value of 0. Rather, we compare models by their accuracy and loss in Tab.\ref{tab2}. 

\begin{table}[h]
\centering
\begin{tabular}{llll}
\hline 
\textbf{Model} &\textbf{Architectural Components} &\textbf{Loss} &\textbf{Accuracy}  \\
\hline 
1: $\eta=0.1$ &Optimizer: Adam &0.4996 &0.5000 \\
 &Optimizer: Gradient Descent &0.4999 &0.5000 \\
 &Optimizer: RMSProp &0.5000 &0.5000 \\
\hline 
2: $\eta=0.01$ &Optimizer: Adam &0.4724 &0.4362 \\
\hline 
3: $\eta=0.001$ &Optimizer: Adam &0.4962 &0.4713 \\
\hline 
4: $\eta=0.001$ &Optimizer: COBYLA &0.4981 &0.4285 \\
\hline 
5: $\eta=0.1$ &Optimizer: Adam &0.5801 &0.4167 \\
 &$p=0.05$ & & \\
 &$\gamma=0.02$ & & \\
\hline 
\end{tabular}
\caption{Comparison of different QNN models.}
\label{tab2}
\end{table}

The depolarizing noise represents the loss of quantum coherence by replacing the quantum state with a mixed state. For a single qubit, with a probability $p$, the state becomes completely mixed, and the density matrix becomes
\begin{equation}
\rho\to\left(1-p\right)\rho+\frac{1}{2}p I,
\end{equation}
where $\rho$ is the density matrix, $p$ is the depolarizing probability, and $I$ is the identity matrix. This noise applies equally to all directions in the Bloch sphere, mimicking errors from imperfect gate implementations or environmental interactions.

Amplitude damping models energy loss in a system, such as a qubit decaying from the excited state ($\ket{1}$) to a ground state ($\ket{0}$). For a single qubit, the Krauss operators are given by
\begin{equation}
E_{0}=
\begin{pmatrix}
1 &0 \\
0 &\sqrt{1-\gamma}
\end{pmatrix}, \quad
E_{1}=
\begin{pmatrix}
0 &\sqrt{\gamma} \\
0 &0 
\end{pmatrix},
\end{equation}
where $\gamma:\ket{1}\to\ket{0}$ is the probability of relaxation, i.e., the probability of an excited state decaying to the ground state. 

Based on the QNN model, we can observe that the Adam optimizer model had the smallest loss and highest accuracy (models that used gradient descent and RMSprop had equal accuracy but reported a higher loss). However, the QNN approach is inherently flawed because all other metrics are 0. Thus, the raw data alone is not sufficient to build an effective QNN, and therefore, feature engineering would be required if one were to adopt the QNN approach effectively.

It is worth noting that if more complexity is introduced into the model, for example, by increasing the number of hidden layers in the network, we might get better performance. However, we speculate that the enhancement in performance will be marginal, and for practical purposes, the QSVC would be the best choice. 


\section{Conclusion}\label{conclusion}

We observe that for pragmatic considerations, the QSVC is the best approach for this dataset because it is easy to implement, has good model metrics, and produces comparatively high accuracy with minimum effort. Unlike some studies that benchmark QML models against classical ML models, we believe that such a comparison is unwarranted and comparable to the age-old adage of comparing ``apples to bananas.'', and thus we avoided building classical ML models to compare; we have compared quantum models with quantum models. 

From the perspective of QML, in the future, we will explore different models and fine-tune the current models implemented to obtain better performance.

Further, in future iterations, we will consider a larger study area to collect more data points and different choices of class-balancing techniques to ascertain which gives optimal performance. In addition to only considering recreation purposes, new models will be built for drinking water quality.

Lastly, we will consider integrating geographical weighting into the existing QSVC and QNN models, for example, like the work in \cite{khan2024cybercrime}, thereby creating a new class of models and proving some results where we show that geographically weighted QNNs have smaller errors, in classifications tasks than standard QNNs.  

\section*{Acknowledgments}
M.A.Z.K. would like to thank Nirvasha Rajdeo for introducing him to this data, checking this research's technical environmental sciences component, and uMngeni-uThukela for granting him permission to use the data. M.AZ.K. and J.A.K. acknowledge that this work is supported by grant number 23070 provided by Zayed University and the Government of the UAE.


\section*{Declarations}

\begin{itemize}
\item \textbf{Funding:} J.A.K. and M.A.Z.K. acknowledge that this research is supported by grant number 23070, provided by Zayed University and the government of the UAE.
\item \textbf{Conflict of interest/Competing interests:} The authors declare that there are no conflicts of interest. 
\item \textbf{Ethics approval and consent to participate:} None required.
\item \textbf{Consent for publication:} The authors grant full consent to the journal to publish this article.
\item \textbf{Data availability:} N/A
\item \textbf{Materials availability:} N/A
\item \textbf{Code availability:} N/A  
\item \textbf{Author contribution:} All authors have contributed equally to this research.
\end{itemize}



\end{document}